\documentclass[apl,amsmath,amssymb,amnsfonts,twocolumn,a4paper]{revtex4-1}
\usepackage[T1]{fontenc}
\usepackage{geometry}
\usepackage{graphicx}
\usepackage{color}
\usepackage[usenames,dvipsnames]{pstricks}
\usepackage{epsfig}

\usepackage{setspace}
\usepackage{booktabs}
\usepackage[percent]{overpic}
\usepackage[strict]{changepage}

\usepackage{todonotes}

\begin{document}

\title{Areal density optimizations for heat-assisted-magnetic recording of high density bit-patterned media} 

\author{Christoph Vogler}
\email{christoph.vogler@tuwien.ac.at}
\affiliation{Institute of Solid State Physics, TU Wien, Wiedner Hauptstrasse 8-10, 1040 Vienna, Austria}
\affiliation{Institute of Analysis and Scientific Computing, TU Wien, Wiedner Hauptstrasse 8-10, 1040 Vienna, Austria}

\author{Claas Abert}
\author{Florian Bruckner}
\author{Dieter Suess}
\affiliation{Christian Doppler Laboratory for Advanced Magnetic Sensing and Materials, Institute for Solid State Physics, TU Wien, Wiedner Hauptstrasse 8-10, 1040 Vienna, Austria}

\author{Dirk Praetorius}
\affiliation{Institute of Analysis and Scientific Computing, TU Wien, Wiedner Hauptstrasse 8-10, 1040 Vienna, Austria}

\date{\today}

\begin{abstract}
Heat-assisted-magnetic recording (HAMR) is hoped to be the future recording technique for high density storage devices. Nevertheless, there exist several realizations strategies. With a coarse-grained Landau-Lifshitz-Bloch (LLB) model we investigate in detail benefits and disadvantages of continuous and pulsed laser spot recording of shingled and conventional bit-patterned media. Additionally we compare single phase grains and bits having a bilayer structure with graded Curie temperature, consisting of a hard magnetic layer with high $T_{\mathrm{C}}$ and a soft magnetic one with low $T_{\mathrm{C}}$, respectively. To describe the whole write process as realistic as possible a distribution of the grain sizes and Curie temperatures, a displacement jitter of the head and the bit positions are considered. For all these cases we calculate bit error rates of various grain patterns, temperatures and write head positions to optimize the achievable areal storage density. Within our analysis shingled HAMR with a continuous laser pulse moving over the medium reaches the best results, and thus having the highest potential to become the next generation storage device.  
\end{abstract}

\maketitle 

\section{Introduction}
\label{sec:intro}
Flash memories are up to become a serious competitor of magnetic storage devices like hard disk drives (HDDs), at least for personal use. While flash drives have their upper hand in speed, HDDs show benefits in reliability, capacity and costs. To be able to maintain this lead the areal storage density (AD) has to increase further, by decreasing the size of each storage bit. Small magnetic grains must have high magnetic anisotropy to ensure high thermal stability of the stored binary information. The anisotropy is limited by the maximum available magnetic field of write heads. One strategy to overcome this so called magnetic recording trilemma is heat-assisted magnetic recording (HAMR) \cite{mee_proposed_1967,guisinger_thermomagnetic_1971,kobayashi_thermomagnetic_1984,rottmayer_heat-assisted_2006}. A laser spot locally heats recording bits near or above the Curie temperature. Due to the resulting reduction of the switching field even the magnetization of very hard magnetic materials can be reversed with available magnetic write fields. A high temperature gradient is usually believed to allow for narrow bits in HAMR, due to a high effective head field gradient
\begin{equation}
\label{eq:write_field}
 \frac{dH}{dx}=\frac{d|H_{\mathrm{ext}}|}{dx}-\frac{dH_{\mathrm{k}}}{dT} \frac{dT}{dx}. 
\end{equation}
The effective head field gradient is believed to be a factor 3-20 larger than in conventional non-thermal recording \cite{kryder_heat_2008}. Nevertheless, thermally written-in errors are a serious problem of HAMR \cite{richter_thermodynamic_2012}, and thus a detailed analysis of switching probabilities and bit error rates (BER) is desirable. In this paper we show how thermal fluctuations at high temperatures during writing, distributions of the Curie temperature of the involved grains and different recording techniques influence the achievable AD and bit transitions. 

A realistic model of the whole write process, including the accurate description of the grain's magnetization dynamics at elevated temperatures, is presented to gain insights into the basic mechanisms of HAMR. The incorporation of temperature in the equation of motion of a magnetic system is computationally challenging, because one has to solve a stochastic partial differential equation with random fluctuations. This requires very short time steps during the integration, and thus even short trajectories are very time consuming. The Landau-Lifshitz-Bloch (LLB) equation \cite{garanin_fokker-planck_1997} allows a coarse spatial discretization of the magnetic grain model. The resulting speed up of the calculations makes it possible to investigate the switching behavior and BER of magnetic grains with realistic dimensions in detail and to estimate how far the areal storage density of bit-patterned media can be increased.

The structure of the paper is as follows: In Sec.~\ref{sec:HAMR_techniques} the examined HAMR techniques are introduced. Section~\ref{sec:models} summarizes the implementation of the LLB equation and explains the coarse grained LLB approach we use in our simulations. Section~\ref{sec:results} presents the resulting switching probability phase diagrams for different materials and HAMR techniques. In this section also the BER for selected grain models calculated from the switching probabilities is shown. Finally areal storage densities are estimated and transition jitters are computed. A conclusion of the results, including recommendations for the design of an optimal HAMR device with high AD, can be found in Sec.~\ref{sec:conclusion}.

\section{HAMR techniques}
\label{sec:HAMR_techniques}
Currently two different strategies how a heat pulse can be applied to a recording medium during a write process are under discussion. In the first approach a pulsed laser spot is moved over the medium. The heat source is just switched on if the write head is located at the correct position to write a grain. The second technique uses a continuous laser spot. We will refer to the two different strategies as PLSR (pulsed laser spot recording) and CLSR (continuous laser spot recording). In the following we illustrate how PLSR and CLSR are considered in our simulations. Both have in common that the applied laser spot is assumed to have a Gaussian spatial distribution with a full width at half maximum (FWHM) of 20\,nm. The way how the time dependent temperature profile at the recording bits and the applied magnetic field are considered, differs for the methods.
\subsection{PLSR}
\label{subsec:PLSR}
During PLSR the movement speed of the write head is assumed to be slow enough that the heat spot can be considered as frozen during the write process of a recording grain. This approximation holds as long as the application time of the heat pulse is short compared to the time the write head needs to cover the distance between two bits. Hence, the heat pulse can be described per:
\begin{eqnarray}
\label{eq:gauss_profile_PLSR}
 T(x,y,t)=T_{\mathrm{min}}&+&\left ( T_{\mathrm{write}}-T_{\mathrm{min}} \right )\cdot\nonumber \\ 
 &&e^{-\frac{\left (x-x_0   \right )^2+\left (y-y_0   \right )^2}{2\sigma^2}-\frac{\left (t-t_0 \right )^2}{\tau^2}},
\end{eqnarray}
with
\begin{equation}
 \sigma=\frac{\mathrm{FWHM}}{\sqrt{8\ln(2)}},\nonumber
\end{equation}
where $x-x_0$ and $y-y_0$ are the down-track and off-track distances between heat spot and bit, respectively. The duration, for which the heat spot is switched on, is denoted with $\tau$, which is typically in the range of $0.1\mathrm{\,ns}<\tau<0.2$\,ns. For a write temperature, at the center of the heat spot, of $T_{\mathrm{write}}$, the pulse reaches the peak temperature $T_{\mathrm{peak}}$ at the recording grain, depending on the relative position to the spot per:
\begin{equation}
 T_{\mathrm{peak}}=\left ( T_{\mathrm{write}}-T_{\mathrm{min}} \right )e^{-\frac{\left (x-x_0   \right )^2+\left (y-y_0   \right )^2}{2\sigma^2}}+T_{\mathrm{min}}.
\end{equation}

A magnetic field is required to align the magnetization of the recording grains in order to store binary information. Since the laser pulse is typically shorter than the magnetic field pulse the latter is assumed as constant during the whole write process in our simulations. 
\subsection{CLSR}
\label{subsec:CLSR}
During CLSR the movement of the heat spot cannot be neglected any more, because the laser is continuously switched on. Each bit along the same track is subject to the same heat pulse with the same peak temperature. $T_{\mathrm{peak}}$ merely decreases in off-track direction. The used temperature profile can be described with:
\begin{equation}
\label{eq:gauss_profile_CLSR}
 T(x,y,t)=\left ( T_{\mathrm{write}}-T_{\mathrm{min}} \right )e^{-\frac{\left (x-vt   \right )^2+\left (y-y_0   \right )^2}{2\sigma^2}}+T_{\mathrm{min}},
\end{equation}
being
\begin{equation}
\label{eq:peak_temp_CLSR}
 T_{\mathrm{peak}}=\left ( T_{\mathrm{write}}-T_{\mathrm{min}} \right )e^{-\frac{\left (y-y_0   \right )^2}{2\sigma^2}}+T_{\mathrm{min}}.
\end{equation}
Here, $v$ is the speed of the write head with a down-track position of $x_0=vt$. $x$ and $y$ denote the down-track and off-track positions of the grains, respectively.

In contrast to PLSR, an applied magnetic field cannot be assumed as constant in time during CLSR any more. As a consequence the correct timing of the field pulse is important to correctly write information on the medium. We model the applied magnetic field with a simple trapezoidal pulse with a field rise and decay time of 0.1\,ns, respectively and a duration of 1.0\,ns. Initially the field points in the $+z$ direction, symmetric with respect to the half simulation time it switches to the $-z$ direction and then back to the $+z$ direction. More precisely, the field direction spans an angle of 0.1\,rad with the easy axis ($z$ direction) of the grains. The down-track position $x$ and the off-track distance $y-y_0$, and thus $T_{\mathrm{peak}}$,  are important for a successful write process. \newline

All performed simulations (PLSR and CLSR) start with an initial temperature of $T_{\mathrm{min}}=270$\,K. The external magnetic field is assumed to be homogeneous in space.
\subsection{shingled vs conventional recording}
\label{subsec:shingledVSconventional}
Besides the above HAMR techniques we distinguish in the following between two established write schemes, shingled and conventional recording. During conventional recording one track is recorded.
\begin{figure}[!h]
\centering
\includegraphics[width=1.15\linewidth]{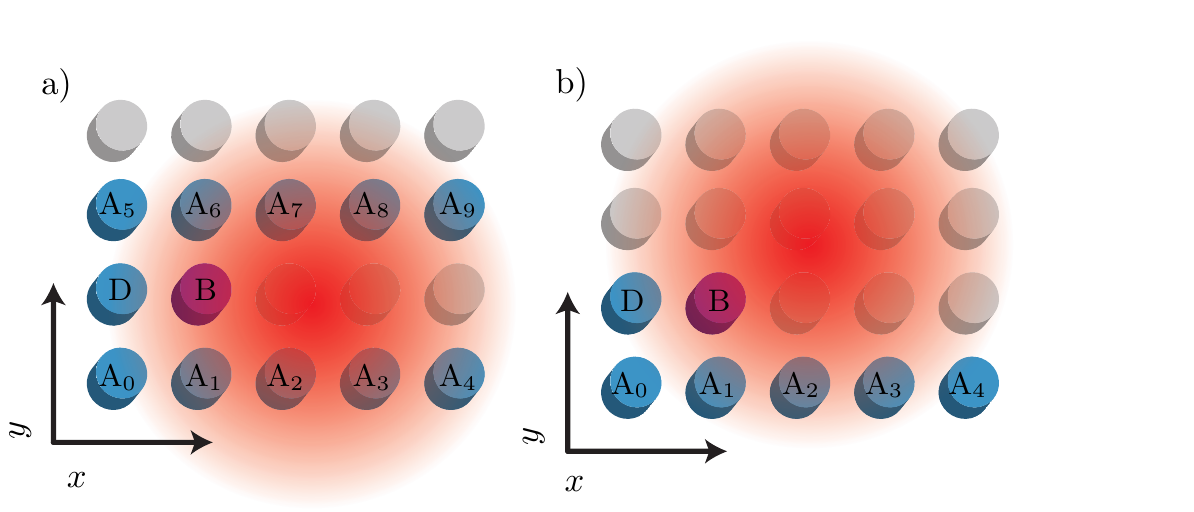}
    \caption{\small (color online) Schematic of the bits involved during a) shingled and b) conventional HAMR. B denotes the bit that should be written. The down-track bit D and all adjacent bits A$_i$ must remain in their original state.}
  \label{fig:writer_pattern}
\end{figure}
Figure~\ref{fig:writer_pattern}a illustrates the situation in which bit B is written. To prevent written-in errors the down-track bit D and all adjacent bits A$_i$ must remain in their original state. As a requirement the adjacent tracks must retain their state for at least 1000 write processes of the center track.

In contrary, during shingled recording (see Fig.~\ref{fig:writer_pattern}b) a whole block of bits is written at once. Hence, solely one adjacent track must be considered in the calculations. The bits A$_i$ must remain unchanged for merely 1 write process.

\section{Coarse grained Landau-Lifshitz-Bloch model}
\label{sec:models}
The magnetization of a magnetic recording grain undergoes a phase transition from a ferromagnetic to paramagnetic state at the Curie temperature  $T_{\mathrm{C}}$. During HAMR the temperature of the applied laser pulse can easily exceed $T_{\mathrm{C}}$. Hence, an accurate simulation model must reproduce this phase transition in each computational cell. The Landau-Lifshitz-Bloch (LLB) equation can fulfill this requirement, as has been confirmed in various publications~\cite{garanin_thermal_2004,chubykalo-fesenko_dynamic_2006,atxitia_micromagnetic_2007,kazantseva_towards_2008,chubykalo-fesenko_dynamic_2006,schieback_temperature_2009,bunce_laser-induced_2010,evans_stochastic_2012,mcdaniel_application_2012,greaves_magnetization_2012,mendil_speed_2013}. Our model is based on the following formulation~\cite{evans_stochastic_2012}:
\begin{eqnarray}
\label{eq:LLB}
  \frac{d \boldsymbol{m}}{dt}= &-&\mu_0{\gamma'}\left( \boldsymbol{m}\times \boldsymbol{H}_{\mathrm{eff}}\right) \nonumber \\
  &-&\frac{\alpha_\perp\mu_0 {\gamma'}}{m^2} \left \{ \boldsymbol{m}\times \left [ \boldsymbol{m}\times \left (\boldsymbol{H}_{\mathrm{eff}}+\boldsymbol{\xi}_{\perp}  \right ) \right ] \right \}\nonumber \\
  &+&\frac{\alpha_\parallel  \mu_0{\gamma'}}{m^2}\boldsymbol{m}\left (\boldsymbol{m}\cdot\boldsymbol{H}_{\mathrm{eff}}  \right )+\boldsymbol{\xi}_{\parallel}.
\end{eqnarray}
Here, $\gamma'$ is the reduced electron gyromagnetic ratio ($\gamma'=|\gamma_{\mathrm{e}}|/(1+\lambda^2)$ with $|\gamma_{\mathrm{e}}|=1.76086\cdot10^{11}$\,(Ts)$^{-1}$), $\mu_0$ is the vacuum permeability and $\boldsymbol{m}$ is the reduced magnetization $\boldsymbol{M}/M_0$, with the saturation magnetization at zero temperature $M_0$. The first two terms on the right-hand side of Eq.~\ref{eq:LLB} describe precession and damping of the system (perpendicular relaxation) and the last two terms denote the length change of the magnetization (longitudinal relaxation) during the time evolution. $\alpha_\parallel$ and $\alpha_\perp$ are dimensionless temperature dependent longitudinal and transverse damping parameters
\begin{equation}
 \alpha_\perp=\begin{cases}\lambda\left( 1-\frac{T}{T_{\mathrm{C}}} \right) & T<T_{\mathrm{C}}\\ \alpha_\parallel & T\geq T_{\mathrm{C}}\end{cases},\quad\alpha_\parallel=\lambda \frac{2T}{3T_{\mathrm{C}}},
\end{equation}
which are connected to the atomistic coupling of the spins to a heat bath with the parameter $\lambda$. To account for thermal fluctuations longitudinal and perpendicular thermal fields $\boldsymbol{\xi}_{\parallel}$ and $\boldsymbol{\xi}_{\perp}$, consisting of white noise random numbers, are used. The main contributions of the effective magnetic field $\boldsymbol{H}_{\mathrm{eff}}$ are the external magnetic field $\boldsymbol{H}_{\mathrm{ext}}$, the anisotropy field
\begin{equation}
  \label{eq:Hani}
   \boldsymbol{H}_\mathrm{ani}=\frac{1}{\widetilde{\chi}_{\perp}(T)}\left( m_x\boldsymbol{e}_{x}+m_y\boldsymbol{e}_{y}\right),
\end{equation}
which is assumed to point along the $z$-direction, and the internal exchange field 
\begin{equation}
\label{eq:blochField}
 \boldsymbol{H}_{\mathrm{J}}=\begin{cases} \frac{1}{2\widetilde{\chi}_{\parallel}(T)}\left( 1-\frac{m^2}{m^2_{\mathrm{e}}(T)} \right)\boldsymbol{m} & T\lesssim T_{\mathrm{C}}\\ -\frac{1}{\widetilde{\chi}_{\parallel}(T)} \left( 1+\frac{3}{5}\frac{T_{\mathrm{C}}}{T-T_{\mathrm{C}}}m^2 \right)\boldsymbol{m}& T\gtrsim T_{\mathrm{C}}.\end{cases}
\end{equation}
Three temperature dependent material functions appear in $\boldsymbol{H}_{\mathrm{ani}}$ and $\boldsymbol{H}_{\mathrm{J}}$, namely the zero field equilibrium magnetization $m_{\mathrm{e}}(T)$, the longitudinal and the perpendicular susceptibility $\widetilde{\chi}_{\parallel}(T)$ and $\widetilde{\chi}_{\perp}(T)$. It has to be mentioned that magnetostatic interactions are not included in the model. To be able to integrate the LLB equation (Eq.~\ref{eq:LLB}) at arbitrary temperatures these functions must be determined for each material which is involved in the particles. We obtain the material functions from simulations with the atomistic code VAMPIRE~\cite{evans_atomistic_2014}, which solves the stochastic Landau-Lifshitz-Gilbert equation for the considered magnetic grains having an atomistic spatial discretization. 

\begin{figure}[!h]
  \centering
  \begin{minipage}{\linewidth}
  \begin{minipage}{0.45\linewidth}
  \includegraphics{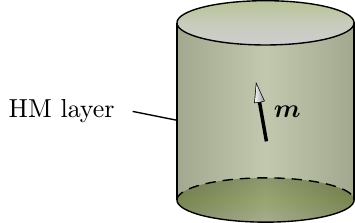}
  \end{minipage}
  \hspace{0.05\linewidth}
  \begin{minipage}{0.45\linewidth}
  \includegraphics{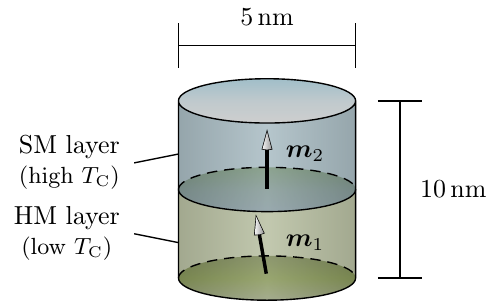}
    \vspace{0.25cm}
  \end{minipage}
  \end{minipage}
    \caption{\small (color online) Monolayer (left) and bilayer (right) grain models. The composite structure on the right consists of a HM layer with low $T_{\mathrm{C}}$ coupled to a SM layer with high $T_{\mathrm{C}}$. Each material layer is represented by a single magnetization vector $\boldsymbol{m}$.}
  \label{fig:layer_model}
\end{figure}

In contrast to an atomistic model, in which the time integration of all magnetic moments in a recording grain, with realistic lateral dimensions of several nanometers, is computationally very expensive, the coarse grained LLB model describes each material with just one vector (see Fig.~\ref{fig:layer_model}). In multilayer structures an intergrain exchange field couples neighboring layers.
Hence, realistic recording grains can be efficiently simulated with low computational effort. Nevertheless, the resulting dynamic trajectories of the atomistic and coarse-grained model are in good agreement~\cite{volger_llb}.

In this work we investigate two different grains for bit-patterned recording. Both have a cylindrical basal plane with a diameter of 5\,nm and a height of 10\,nm. We compare a single hard magnetic (HM1) grain and a bilayer structure with graded Curie temperature, consisting of a HM2 material layer coupled to a soft magnetic (SM) one in a ratio of 50:50 (see Fig.~\ref{fig:layer_model}). Both HM layers have large uniaxial anisotropy. They differ in their Curie temperature. The SM material is a perfect soft magnet with a large saturation magnetization comparable to that of pure Fe. Table~\ref{tab:mat_match} illustrates the detailed material constants.
\begin{table}[h!]
  \centering
  \vspace{0.5cm}
  \begin{tabular}{c c c c}
    \toprule
    \toprule
      & HM1 & HM2 & SM \\
    \midrule
    $K_1$\,[J/m$^3$] & $6.6\cdot 10^6$ &  $6.6\cdot 10^6$ & 0.0\\
    $J_{\mathrm{S}}$\,[T] & 1.43 & 1.43 & 2.16 \\
    $\lambda$ & 0.1 & 0.1 & 1.0\\
    $T_{\mathrm{C}}$\,[K] & 536.94 & 744.12 & 1140.67\\
    \bottomrule
    \bottomrule
  \end{tabular}
  \caption{\small Material parameters of a soft magnetic (SM) two hard magnetic (HM1 \& HM2) layers. $K_1$ is the anisotropy constant, $J_{\mathrm{S}}$ the saturation magnetization, and $\lambda$ the damping constant. }
  \label{tab:mat_match}
\end{table}
\subsection{magnetostatic interactions}
\label{subsec:interactions}
\begin{figure}
  \centering
  \includegraphics{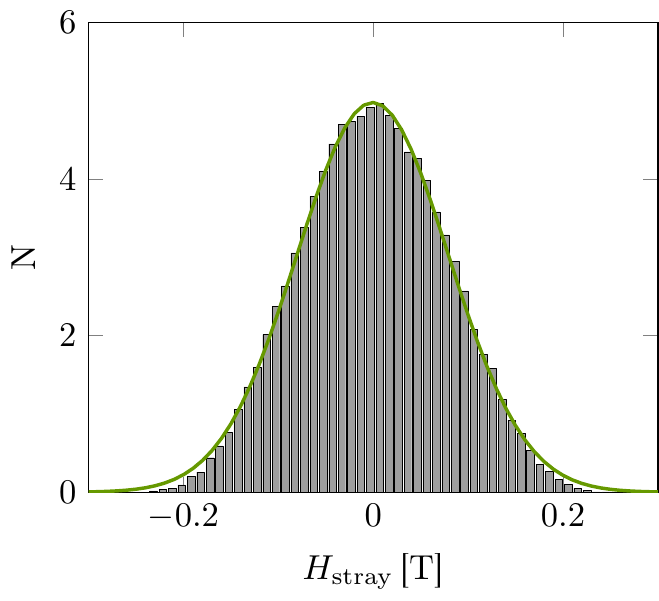}
  \caption{\small (color online) Histogram of total strayfields $H_{\mathrm{stray}}$ at the center grain for 50\,000 different magnetization configurations of 24 surrounding grains. The center to center down-track and off-track spacing between the grains are $l_x=7.5$\,nm and $l_y=6.0$\,nm, respectively. The solid green line shows a fit with a standard normal distribution.}
  \label{fig:histogram}
\end{figure}

In the presented model of Sec.~\ref{sec:models} no interactions between grains are included in the equation of motion (Eq.~\ref{eq:LLB}). To estimate the influence of the strayfield of neighboring grains on the center bit we consider a block of 5x5 grains in a full micromagnetic model. All grains have the same size (5\,nm diameter and 10\,nm length) and material. To compute the distribution of the produced magnetic field of all 24 surrounding grains at their center, the field for 50\,000 different magnetization configurations is simulated. The magnetization direction of each grain (up or down) is randomly chosen in each configuration. A histogram of the resulting field values is then fitted with a standard normal distribution. This allows to extract the standard deviation of the field distribution. In the case of pure HM grains the histogram for a block with a center to center down-track and off-track spacing of $l_x=7.5$\,nm and $l_y=6.0$\,nm, respectively, is shown in Fig.~\ref{fig:histogram}.
In the worst case all 24 surrounding grains have parallel magnetization and produce a magnetic field, which tries to align the center bit in an antiparallel direction. Figure~\ref{tab:field_dist} summarizes the results for single phase and bilayer grains for various bit patterns.
\begin{table}
  \centering
  \vspace{0.5cm}
  \begin{tabular}{c c c c c c}
    \toprule
    \toprule
      grain & $l_x$\,[nm] & $l_y$\,[nm] & $\sigma_{H_{\mathrm{stray}}}$\,[T] & $\sigma_{\mathrm{stray}}$\,[\%\,$T_{\mathrm{C}}$] & $\sigma_{\mathrm{T}}$\,[\%\,$T_{\mathrm{C}}$] \\
    \midrule
      HM1 & 7.5 & 6.0 & 0.080 & 0.95 & 3.15\\
      HM2/SM & 7.5 & 6.0 & 0.100 & 5.60 & 6.35 \\
      HM2/SM & 7.5 & 14.0 & 0.054 & 3.00 & 4.25 \\
      HM2/SM & 10.0 & 10.5 & 0.038 & 2.12 & 3.70 \\
      HM2/SM & 12.0 & 9.5 & 0.037 & 2.05 & 3.65\\
    \bottomrule
    \bottomrule
  \end{tabular}
  \caption{\small Distribution of the strayfield at the center produced from 24 neighboring grains. $l_x$ and $l_y$ denote the center to center down-track and off-track spacing, respectively. The strayfield causes an additional contribution $\sigma_{\mathrm{stray}}$ to the intrinsic $T_{\mathrm{C}}$ distribution, resulting in a total distribution $\sigma_{\mathrm{T}}$, as described in Sec.~\ref{subsubsec:effect_Tc_strayfield}.}
  \label{tab:field_dist}
\end{table}

\section{Results}
\label{sec:results}
\subsection{pulsed laser spot recording (PLSR)}
\label{sec:plsr_results}
\begin{figure}[!h]
\begin{adjustwidth}{-1cm}{-1cm}
\includegraphics{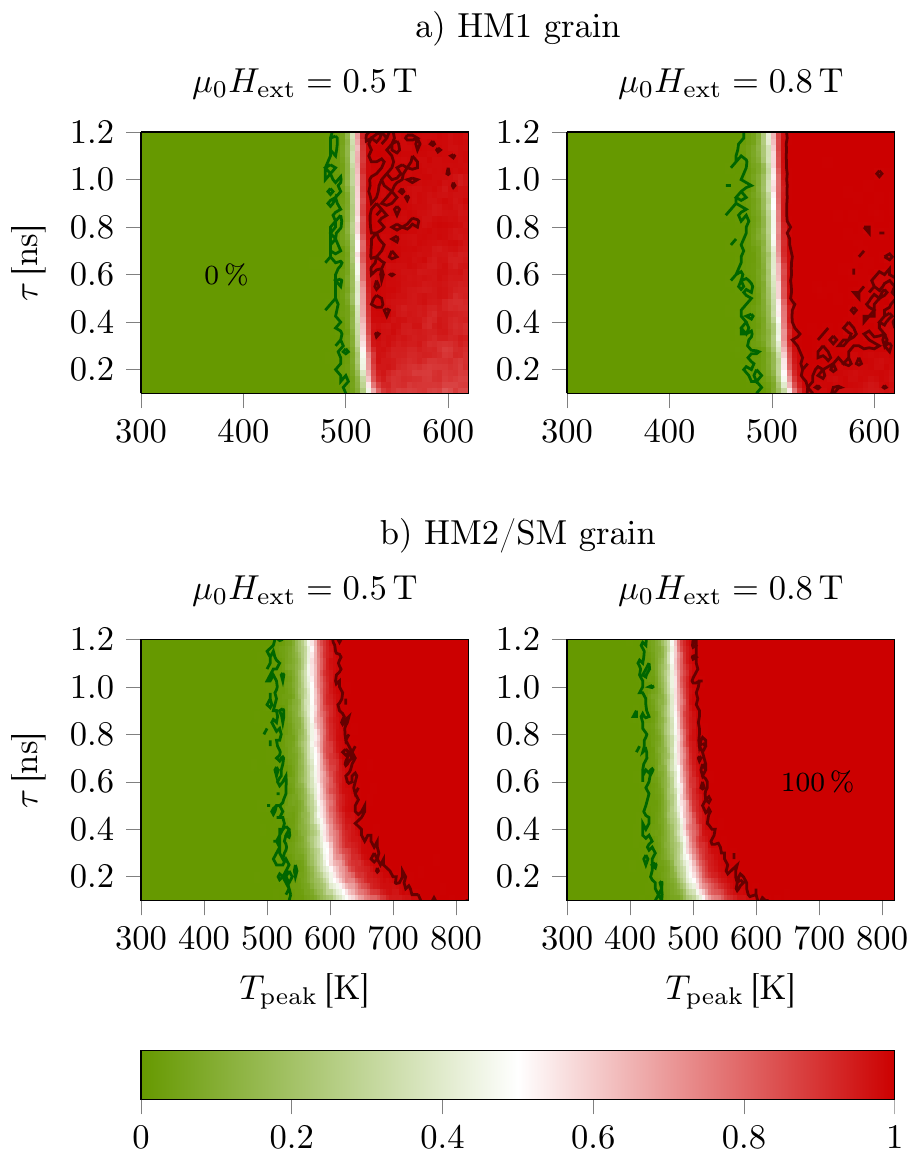}
  \caption{\small (color online) Switching probability phase diagrams for the a) single phase HM1 grain the b) HM2/SM bilayer structure with graded Curie temperature for PLSR. Two different external magnetic fields are applied to the grains. Each phase point consists of 128 switching trajectories with the same heat pulse duration $\tau$ and peak temperature $T_{\mathrm{peak}}$. The solid lines mark the phase transitions between the dark red areas, where complete switching (all 128 trajectories switched) occurs and the light green areas, where no switching is possible.}
  \label{fig:plsr_phase}
\end{adjustwidth}
\end{figure}
In this section we analyze the recording performance of PLSR on bit-patterned media. Our goal is to gain insights into the switching behavior of the presented grains to optimize the achievable areal storage density (AD). Since the coarse grained LLB approach is fast and reliable several phase diagrams of the switching probability for various external magnetic fields and temperature pulses can be simulated. In each simulation a full heat pulse is applied to the grain with an additional constant external field, which tries to switch the particle from the $+z$ to the $-z$ direction. After the simulation the state of the particle (switched or not) is evaluated. In each phase point 128 trajectories are calculated to obtain a switching probability. Figure~\ref{fig:plsr_phase} presents the resulting switching probability phase diagrams, where the influence of the heat pulse length and its peak temperature is examined. The resolution of the diagram in the temperature axis is $\Delta T=5$\,K and in the pulse duration axis $\Delta \tau=50$\,ps. Hence, each diagram contains the data of about 180 000 switching trajectories. The contour lines mark the phase transition between the dark red areas, where 100\,\% of the trajectories (all of the 128 simulations) switch, and the light green areas, where no single trajectory switches.

To ensure high recording speed typically laser pulse durations of $0.1$\,ns to $0.2$\,ns are used for PLSR. Nevertheless, the scan of the phase space in Fig.~\ref{fig:plsr_phase} is performed up to much lager values of $\tau$. The reason is that for short heat pulses the switching probability is well below 100\,\% in the case of a HM1 monolayer (Fig.~\ref{fig:plsr_phase}a). This holds for both examined external fields of 0.5\,T and 0.8\,T. The switching behavior improves if the duration of the heat pulse increases. By looking at short pulses one observes that the available time for the grain's magnetization reversal under the heat pulse is too short in case of the HM1 grain, in contrast to long pulses. Similar results were obtained for granular media in \cite{zhu2013understanding,zhu2015medium} due to a short effective recording time window. Hence, we examine a HM2/SM composite structure with graded Curie temperature and high damping in the SM part. Figure~\ref{fig:plsr_phase}b points out that due to the faster magnetization relaxation the switching probabilities are similar for short and for long pulse durations. The switching probability reaches 100\,\% for all analyzed values of $\tau$, but the transition area is broader for short pulses. Furthermore, a magnetic field of 0.5\,T is sufficient to switch the grain at high temperatures even for short pulses.

We designed the bilayer structure that the transition temperature is approximately the same as for the monolayer grain. Note that this transition temperature is clearly below the Curie temperature of the involved HM2 and SM materials (compare with Tab.~\ref{tab:mat_match}). In contrast, the transition of the switching probability in case of a pure HM1 grain is just slightly below $T_{\mathrm{C}}$. Hence, the reduction of the switching temperature due to the graded Curie temperature can be clearly seen. 

\subsubsection{effects of $T_{\mathrm{C}}$ distributions and magnetostatic interactions}
\label{subsubsec:effect_Tc_strayfield}
\begin{figure}
 \includegraphics{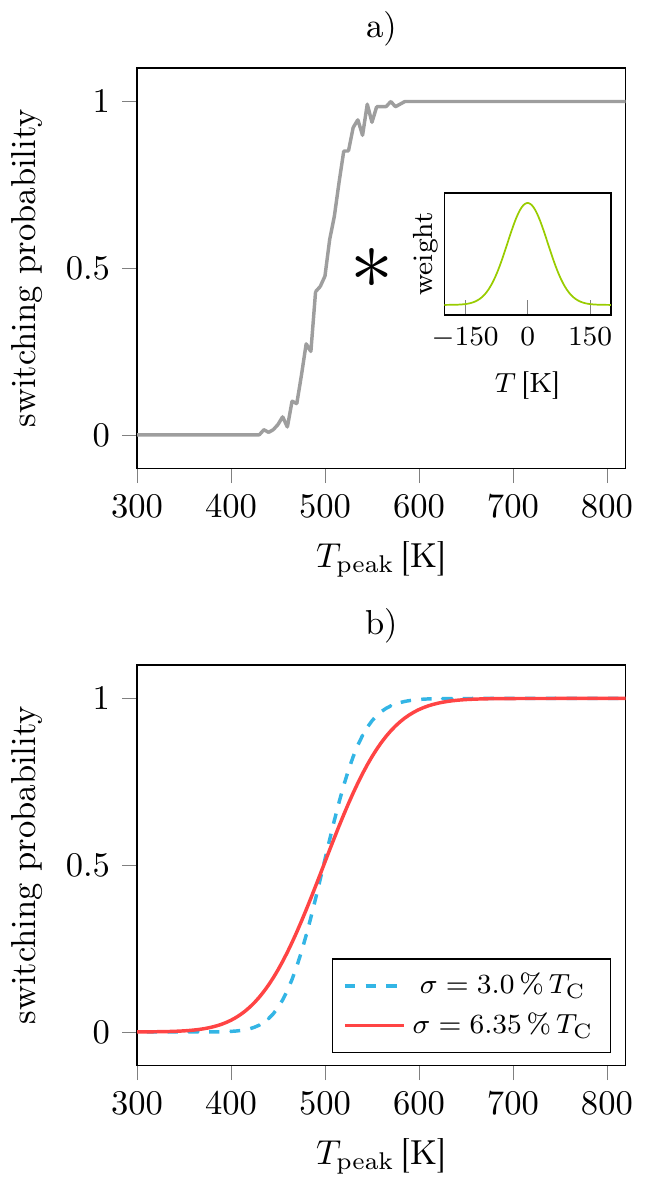}
  \caption{\small Consideration of magnetostatic interactions in terms of an additional contribution to the intrinsic Curie temperature distribution. a) Switching probability of the HM2/SM grain for a pulse duration of $\tau=0.2$\,ns and under an external applied field of 0.8\,T, for various peak temperatures. The inset shows the Gaussian weighting function which is used for the convolution with the original probabilities. b) Resulting switching probabilities for $\sigma_{\mathrm{T}}=3.0\,\%\,T_{\mathrm{C}}$ and $\sigma_{\mathrm{T}}=6.35\,\%\,T_{\mathrm{C}}$.}
  \label{fig:convolution}
\end{figure}

In a real world application recording grains of bit-patterned media never have perfectly equal sizes and material properties. These variations change the Curie temperature of the bits. Hence, a $T_{\mathrm{C}}$ distribution with a standard deviation of typically $3\,\%\,T_{\mathrm{C}}$ must be considered. A shift of $T_{\mathrm{C}}$ results in a shift of the transition of the switching probability. To incorporate $T_{\mathrm{C}}$ distributions in the simulated phase diagrams we perform a convolution of the switching probability for a fixed pulse duration $\tau$ and a Gaussian weighting function with $\sigma_{T_{\mathrm{C}}}=3\,\%\,T_{\mathrm{C}}$. This procedure is illustrated in Fig.~\ref{fig:convolution} for the HM2/SM grain with $\tau=0.2$\,ns and $\mu_0H_{\mathrm{ext}}=0.8$\,T. 

Varying external fields have the same effect of a transition temperature shift (see Fig.~\ref{fig:plsr_phase}). Thus, we consider the magnetostatic interactions between bits (see Sec.~\ref{subsec:interactions}) with a further contribution to the $T_{\mathrm{C}}$ distribution. In the case of a difference in the external field of $\Delta \mu_0 H_{\mathrm{ext}}=0.3$\,T we obtain a total shift of $\Delta T_{\mathrm{peak}}=124$\,K for the HM2/SM grain and a shift of $\Delta T_{\mathrm{peak}}=19$\,K for the pure HM1 grain. With the knowledge of the field distribution the temperature shift can be converted to an additional $T_{\mathrm{C}}$ distribution (see Tab.~\ref{tab:field_dist}). In total a distribution with $\sigma_{\mathrm{T}}=\sqrt{\sigma^2_{T_\mathrm{C}}+\sigma^2_{\mathrm{stray}}}$ must be considered. 

For the same pattern the composite structure shows a larger additional distribution than the pure HM1 grain. This is a consequence of the large saturation magnetizations of the layers, which are not optimal in graded structures \cite{suess_exchange_2005}. The pure HM1 grain has a much smaller temperature shift of the transition. Nevertheless, the HM1 grain cannot be used in an application, because it has no 100\,\% switching probability even at high peak temperatures in case of a pulse duration of $\tau=0.2$\,ns.
 
\subsubsection{bit error rate (BER) and areal density (AD)}
\label{sec:BER_PLSR}
Due to the knowledge of the detailed switching probabilities of the investigated recording grains we can directly calculate the BER of a recording process for a given grain arrangement in bit-patterned media. The BER can be calculated per:
\begin{equation}
\label{eq:BER}
 \mathrm{BER}=1-P_{\mathrm{B}}\left ( 1-P_{\mathrm{D}} \right )\prod_{i=0}^{j}\left ( 1-P_{\mathrm{A}_i} \right )^{n}.
\end{equation}
Here, $P_{\mathrm{B}}$ is the switching probability of bit B (see Fig.~\ref{fig:writer_pattern}), $1-P_{\mathrm{D}}$ the probability that the down-track bit D does not switch and $1-P_{\mathrm{A}_i}$ the probability of not switching the adjacent bit A$_i$. The exponent $n$ in the joint probability considers the number of write processes during which the adjacent bits must retain their magnetic state. As explained in Sec.~\ref{subsec:shingledVSconventional} $n=1000$ holds for conventional recording and $n=1$ for shingled recording. With Eq.~\ref{eq:BER} the BER can be determined for a given
\begin{itemize}
 \item down-track and off-track spacing ($l_x$ and $l_y$) of the recording grains
 \item write temperature of the heat spot $T_{\mathrm{write}}$
 \item down-track and off-track position ($x$ and $y$) of the heat spot.
\end{itemize}
Based on $T_{\mathrm{write}}$ and the distance between heat spot center and recording bit the peak temperature $T_{\mathrm{peak}}$ at each bit can be computed. The corresponding switching probability from Fig.~\ref{fig:plsr_phase} allows the calculation of the BER per Eq.~\ref{eq:BER}. 

\begin{figure}
\begin{adjustwidth}{-1cm}{-1cm}
 \includegraphics{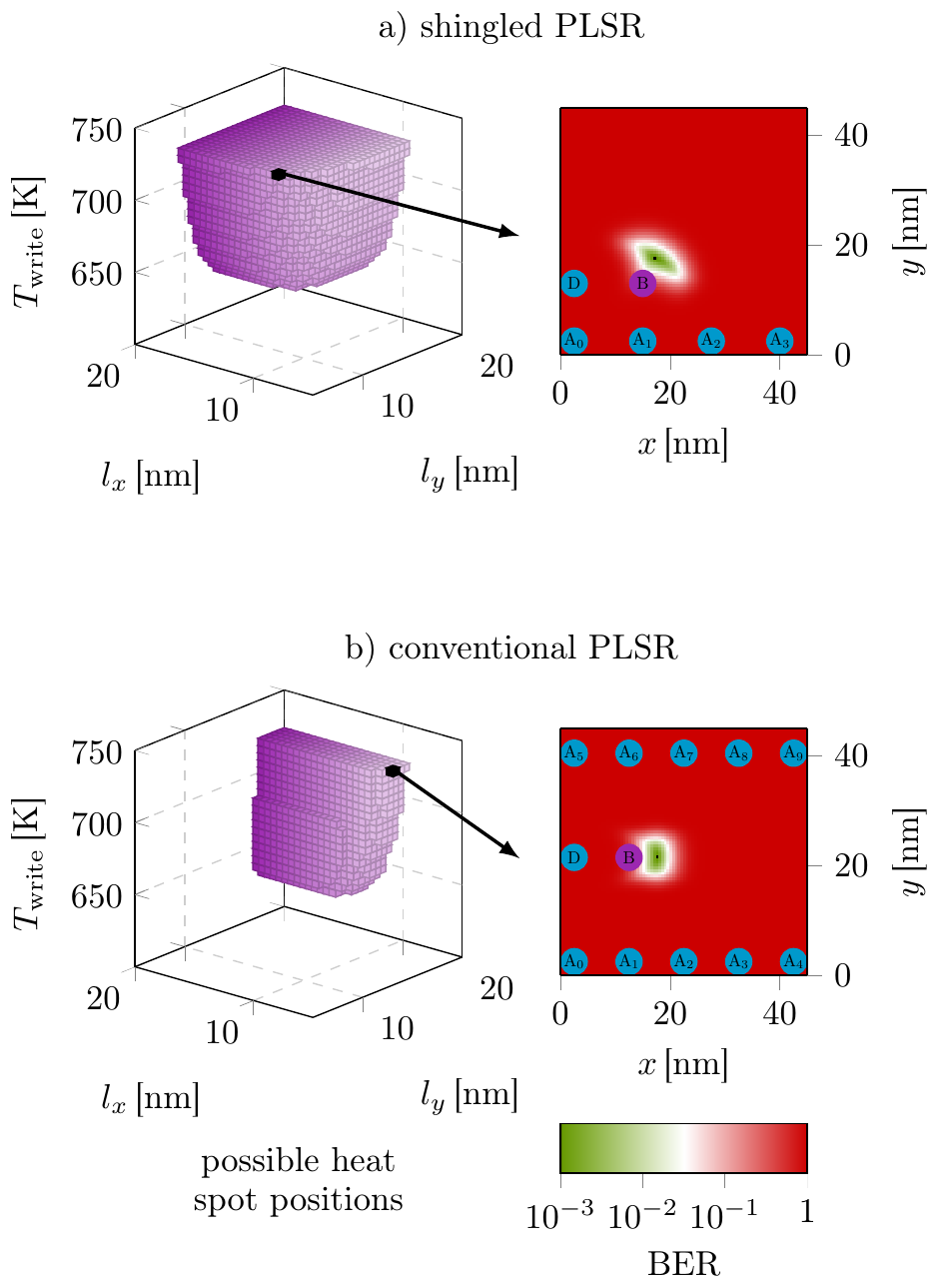}
  \caption{\small (color online) AD optimizations for a) shingled and b) conventional PLSR of a medium consisting of HM2/SM grains, subject to a write field of 0.8\,T.\newline
  (left) On the left a map of possible grain spacings and write temperatures to successfully switch bit B under the constraint of $\text{BER}<10^{-3}$ is shown. A distribution of the Curie temperature of 3.65\,\% and a displacement jitter according to Eqs.~\ref{eq:displacement_jitter} and \ref{eq:displacement_jitter_tot} are considered. A color gradient is used for the purpose of clearer presentation. \newline  
  (right) In the case of the setup with the highest possible AD, the BER for various heat spot positions $x$ and $y$ is shown in detail on the right. The black points display positions where $\text{BER}<10^{-3}$. The bit B should be written while all neighboring bits D and A$_i$ should retain their state.}
  \label{fig:plsr_BER}
\end{adjustwidth}
\end{figure}

We account for grain size and displacement distributions ($\sigma_{\mathrm{bitSize}}$ and $\sigma_{\mathrm{bitPos}}$) as well as a distribution of the write head position ($\sigma_{\mathrm{headPos}}$) per the guidelines of the Advanced Storage Technology Consortium (ASTC):
\begin{eqnarray}
\label{eq:displacement_jitter}
  &\sigma_{\mathrm{bitSize}}&=5\,\%\,\min(l_x,l_y)\nonumber \\
  &\sigma_{\mathrm{bitPos}}&=5\,\%\,\min(l_x,l_y)\nonumber \\
  &\sigma_{\mathrm{headPos}}&=2\,\%\,l_x.
\end{eqnarray}
The total displacement jitter is
\begin{equation}
  \label{eq:displacement_jitter_tot}
  \sigma_{\mathrm{displ}}=\sqrt{\sigma_{\mathrm{bitSize}}^2+\sigma_{\mathrm{bitPos}}^2+\sigma_{\mathrm{headPos}}^2}.
\end{equation}
For fixed $T_{\mathrm{write}}$, $l_y$ and $l_y$ we calculated a BER map for various heat spot positions $x$ and $y$ in a range of 0\,nm to 45\,nm ($\Delta x=\Delta y = 0.5$\,nm) per Eq.~\ref{eq:BER} (as for example illustrated in the right-hand side of Fig.~\ref{fig:plsr_BER}). The resulting map was then corrected with the total displacement jitter by performing a convolution of the BER map and a two-dimensional Gaussian weighting function with the standard deviation being $\sigma_{\mathrm{displ}}$. As a consequence, the write process is counted as successful if at least one heat spot position with $\mathrm{BER}<10^{-3}$ was found. The grain spacings $l_x$ and $l_y$ and the write temperature $T_{\mathrm{write}}$ were then varied, with the objective to maximize the AD.

Figure~\ref{fig:plsr_BER} displays the results of the optimization process in the case of a medium with HM2/SM grains, a pulse duration of $\tau=0.2$\,ns, an external magnetic field of $0.8$\,T and $\sigma_{\mathrm{T}}=3.65$\,\%\,$T_{\mathrm{C}}$. On the left hand side possible values of the grain spacings and the write temperature, which satisfy the demanded constraint of $\text{BER}<10^{-3}$, are plotted with purple cubes (grid: $\Delta l_x = \Delta l_y=0.5$\,nm, $\Delta T_{\mathrm{write}}=5$\,K). On the right hand side the BER map for various heat spot positions is illustrated for the case of a maximum AD (optimal $l_x$, $l_y$ and $T_{\mathrm{write}}$). For shingled PLSR (Fig.~\ref{fig:plsr_BER}a) one notices that the minimum grain spacings in down- and off-track direction decrease for increasing write temperatures. The reason is that the heat spot can be positioned off-track, and thus the higher thermal gradient for higher write temperatures can be used to lower the grain spacings. In contrast, for conventional PLSR (Fig.~\ref{fig:plsr_BER}c) merely the down-track spacing $l_x$ decreases for increasing $T_{\mathrm{write}}$. The off-track spacing $l_y$ seems to remain almost constant, independent from the write temperature. Naively one would expect $l_y$ to decrease for decreasing write temperatures. One cannot position the heat spot off-track. In this picture the adjacent tracks are subject to lower peak temperatures for decreasing $T_{\mathrm{write}}$, yielding lower switching probabilities, and thus lower BER. But, one has to consider that the adjacent tracks have to remain their magnetic state for 1000 write processes. Hence, even low switching probabilities of $P_{\mathrm{A}_i}>10^{-6}$ lead to $\text{BER}>10^{-3}$. As a consequence the adjacent bits must almost cool down to room temperature. The actual temperature at a distance from the heat spot of the FWHM is approximately the same ($\sim300$\,K) for all investigated write temperatures (see Eq.~\ref{eq:gauss_profile_PLSR}). Since the FWHM of the heat pulses are identical for all $T_{\mathrm{write}}$, the effect of Fig.~\ref{fig:plsr_BER}c becomes clear. If $n=1$ would be valid, we would in fact see a slightly decreasing $l_y$ for decreasing $T_{\mathrm{write}}$ for conventional recording, because the above argument does not hold at elevated temperatures, where a high thermal gradient appears.

\begin{table}[h!]
  \centering
  \vspace{0.5cm}
  \begin{tabular}{c c c c c c}
    \toprule
    \toprule
      & $\sigma_{\mathrm{T}}$\,[$\%\,T_{\mathrm{C}}$] & $l_x$\,[nm] & $l_y$\,[nm] &  $T_{\mathrm{write}}$\,[K]& AD\,[Tb/in$^2$] \\
    \midrule
      S & 3.0 & 12.0 & 9.5 & 720.0 & 5.66\\
      C & 3.0 & 9.0 & 17.5 & 720.0 & 4.10\\
    \midrule
      S & 3.65 & 12.5 & 10.5 & 720.0 & 4.92\\
      C & 3.65 & 10.0 & 19.0 & 720.0 & 3.40\\
    \bottomrule
    \bottomrule
  \end{tabular}
  \caption{\small Optimal center to center grain spacings, write temperatures and areal storage densities for shingled (S) and conventional (C) PLSR of HM2/SM bilayers with and without considering magnetostatic interactions. $\mu_0 H_{\mathrm{ext}}$ is assumed to be 0.8\,T. The shown values of $l_x$, $l_y$, $T_{\mathrm{write}}$ and the AD are all outputs of the optimization process.}
  \label{tab:plsr_BER}
\end{table}
Table~\ref{tab:plsr_BER} summarizes the optimized parameters and obtained AD. For shingled PLSR the grain spacings in down- and off-track direction are similar, which is not surprising due to symmetry considerations. For conventional recording the off-track spacing is twice as large as the down-track spacing. This fact is again a result of $n=1000$ in Eq.~\ref{eq:BER}. With shingled PLSR one can reach almost 5\,Tb/in$^2$, which is a significant improvement compared to magnetic recording without heat-assist. No BER calculations were preformed for the pure HM1 grain, because the switching probability does not reach 100\,\% (see Fig.~\ref{fig:plsr_phase}). 

Note that we assumed the magnetostatic interactions to be $\sigma_{\mathrm{stray}}=0.65\%\,T_{\mathrm{C}}$ (according to $l_x=12$\,nm and $l_y=9.5$\,nm in Tab.~\ref{tab:field_dist}). Although the final grain spacings with strayfield are larger (and thus the interactions are actually slightly overestimated) we fix its influence to the case with the densest pattern without $\vec{H}_{\mathrm{stray}}$, which gives an upper bound. In principle one could iteratively calculate the down-track and off-track spacing with the exact magnetostatic interactions, but this is a tedious approach, which just marginally influences the results.

\subsubsection{transition jitter}
\label{subsubsec:jitter_PLSR}
\begin{figure}
\includegraphics{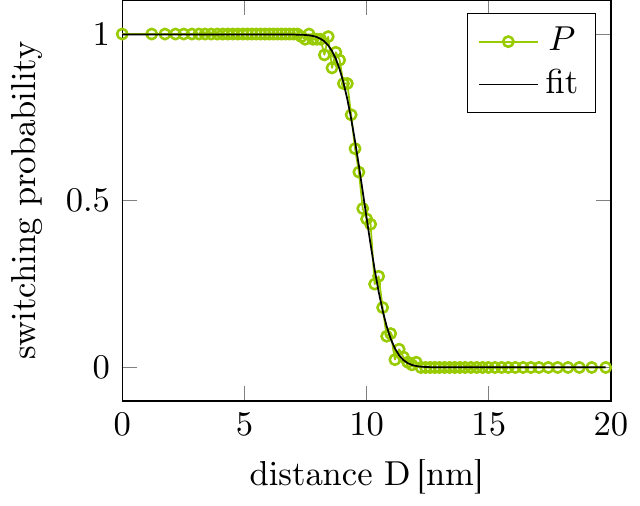}
\caption{\small Switching probability versus distance from the heat spot center with $T_{\mathrm{write}}=720$\,K. A HM2/SM grain with a distribution of the Curie temperature of $\sigma_{\mathrm{T}}=3\%\,T_{\mathrm{C}}$, subject to an external magnetic field of 0.8\,T, is investigated.}
\label{fig:transition_jitter_fit_PLSR}
\end{figure}

If the simulated island is interpreted as one grain of a granular medium, the transition jitter can be computed from the switching probability phase diagrams (see Fig.~\ref{fig:plsr_phase}). With the known dependency of the switching probability on the peak temperature, the optimal write temperature and the known heat pulse shape, $T_{\mathrm{peak}}$ can be transfered to a distance. For example, if we assume $T_{\mathrm{peak}}=600$\,K at some position. We know that the heat pulse has a Gaussian shape with $\text{FWHM}=20$\,nm and a write temperature of 720\,K at the spot center. Hence, the distance between the heat spot center and the considered position can easily be evaluated. This can be done for any peak temperature yielding a relation between switching probability and distance from the heat spot center as illustrated in Fig.~\ref{fig:transition_jitter_fit_PLSR}. Fitting this curve with the cumulative distribution function of a normal distribution yields the desired transition jitter $\sigma_{dP/d\mathrm{D}}$. 
In Tab.~\ref{tab:plsr_transition_jitters} various transition jitters for a HM2/SM grain are listed.
\begin{table}[h!]
  \centering
  \vspace{0.5cm}
  \begin{tabular}{c c c c}
    \toprule
    \toprule
      $\sigma_{\mathrm{T}}$\,[$\%\,T_{\mathrm{C}}$] & $T_{\mathrm{write}}$\,[K] & $\mu_0 H_{\mathrm{ext}}$\,[T] & $\sigma_{dP/d\mathrm{D}}$\,[nm] \\
    \midrule
      0.0 & 720.0 & 0.5 & 1.08\\
      0.0 & 620.0 & 0.8 & 1.00\\
      0.0 & 720.0 & 0.8 & 0.80\\
    \midrule
      3.0 & 720.0 & 0.5 & 1.31\\
      3.0 & 620.0 & 0.8 & 1.36\\
      3.0 & 720.0 & 0.8 & 1.08\\
    \midrule
      3.65 & 720.0 & 0.5 & 1.40\\
      3.65 & 620.0 & 0.8 & 1.50\\
      3.65 & 720.0 & 0.8 & 1.20\\
    \bottomrule
    \bottomrule
  \end{tabular}
  \caption{\small Transition jitters for various Curie temperature distributions, external magnetic fields and write temperatures, in the case of a HM2/SM grain and a pulse duration of $\tau=0.2$\,ns.}
  \label{tab:plsr_transition_jitters}
\end{table}
The transition jitter decreases for an increasing external field, if the same write temperatures are assumed. For decreasing write temperatures $\sigma_{dP/d\mathrm{D}}$ increases, because of the lower thermal gradient. The jitter also increases if $T_{\mathrm{C}}$ distributions and magnetostatic interactions are considered. One notices that the jitter increase due to 3\,\% $T_{\mathrm{C}}$ distribution is similar to its decrease for a 0.3\,T higher magnetic field (compare line 1 and 6 in Tab.~\ref{tab:plsr_transition_jitters}). 
Finally, it has to be mentioned that in a realistic situation with $T_{\mathrm{C}}$ distributions the jitter for PLSR is much larger than for perpendicular recording without heat-assist \cite{suess_superior_2015}

\subsection{continuous laser spot recording (CLSR)}
\label{sec:clsr_results}
\begin{figure}[!h]
\includegraphics{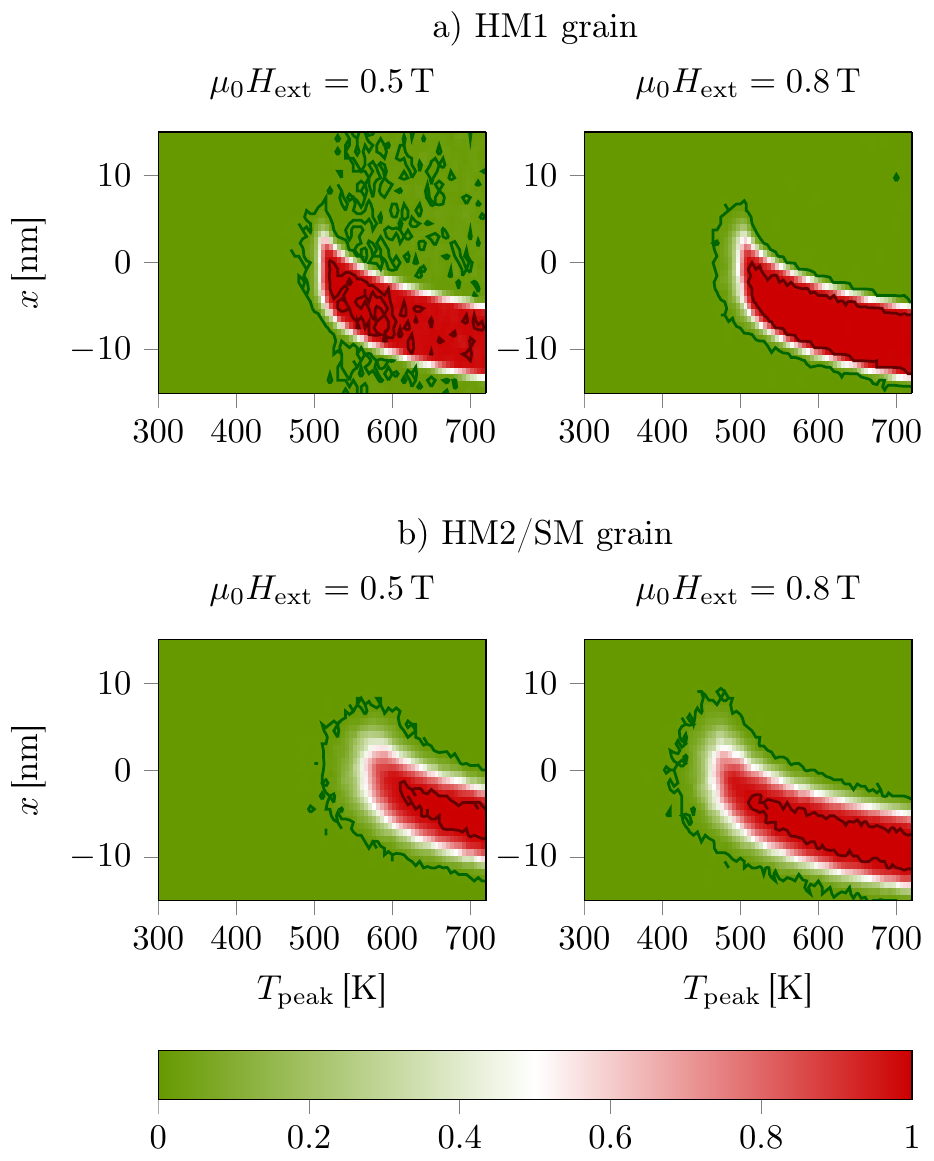}
  \caption{\small (color online) Switching probability phase diagrams for a single phase HM1 grain and a HM2/SM structure with graded Curie temperature for CLSR with a head velocity of 7.5\,m/s. Two different external magnetic fields are applied to the grains. Each phase point consists of 128 switching trajectories with the same down-track position $x$ and the same peak temperature $T_{\mathrm{peak}}$. The color code and the threshold of the solid lines, which mark the phase transition areas, are equivalent to those in Fig.~\ref{fig:plsr_phase}.}
  \label{fig:clsr_phase}
\end{figure}
\begin{figure*}
\includegraphics{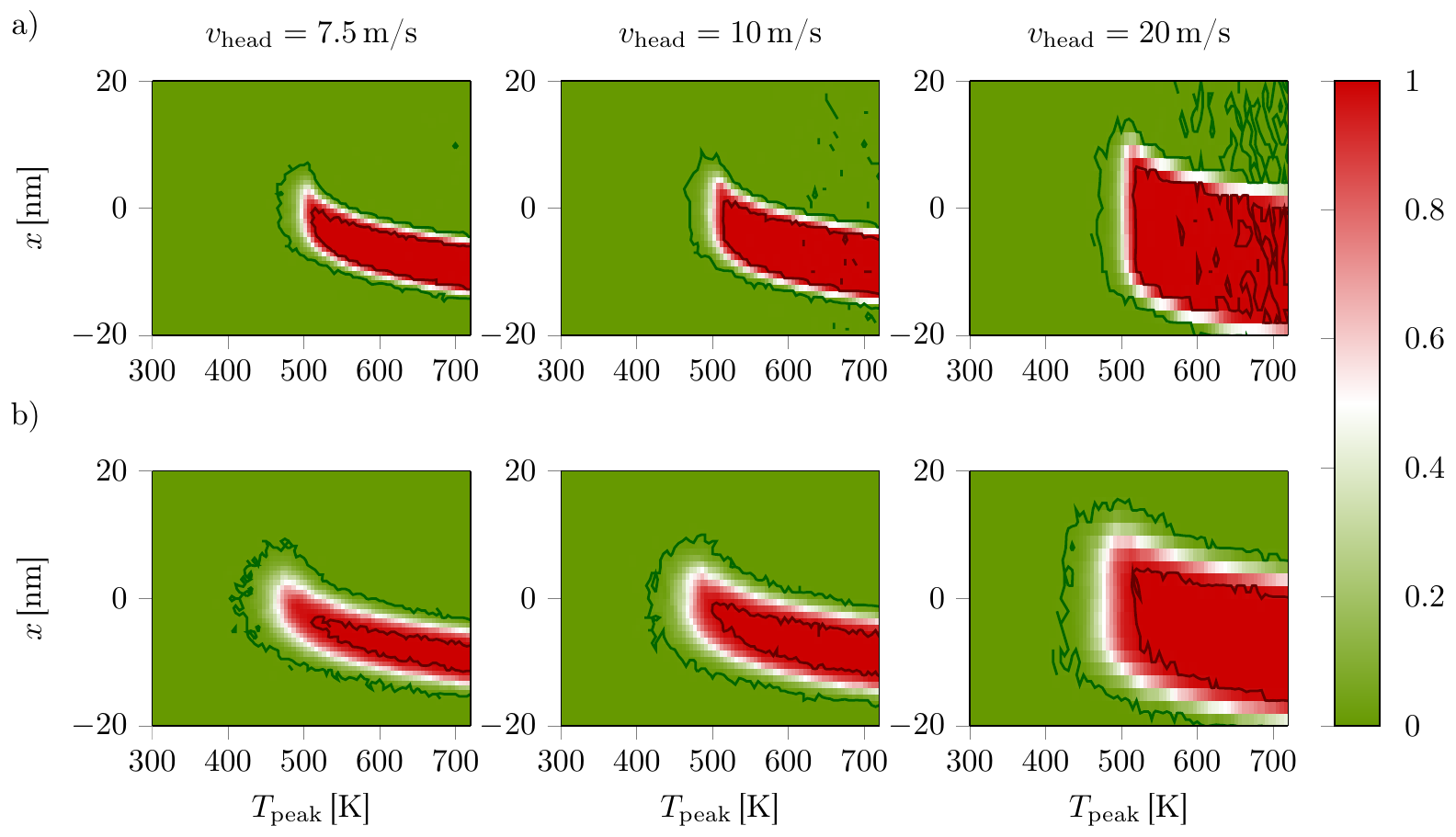}
  \caption{\small (color online) Switching probability phase diagram of the a) HM1 b) HM2/SM grain under an applied field of 0.8\,T. Various head velocities $v_{\mathrm{head}}$ are compared.}
  \label{fig:clsr_vel_comp}	
\end{figure*}
Similar to PLSR we extend the study for CLSR in this section. In contrast to PLSR, a continous heat spot moves over the recording medium, which can no longer be treated as quasi-static. According to Eq.~\ref{eq:gauss_profile_CLSR} the head velocity plays an important role. Since the profile of the external magnetic field is fixed with a duration of 1\,ns in our simulations, the head velocity limits the minimum down-track spacing. Another difference to PLSR is that along one track all grains are subject to the same peak temperature, because the heat spot moves in down-track direction over all bits. $T_{\mathrm{peak}}$ just decreases in off-track direction. As a consequence the down-track position $x$ and the peak temperature $T_{\mathrm{peak}}$ (or rather the off-track distance $y-y_0$ via Eq.~\ref{eq:peak_temp_CLSR}) of the grains are crucial parameters.

Figure~\ref{fig:clsr_phase} presents phase diagrams of the switching probability for the same grain models as in Sec.~\ref{sec:plsr_results}. The resolution in the peak temperature axis is again $\Delta T=5$\,K and in down-track direction a resolution of $\Delta x=0.75$\,ns and a head velocity of $v_{\mathrm{head}}=7.5$\,m/s is used. Hence, each phase diagram contains the data of almost 250 000 switching trajectories with a length of 8.5\,ns. The phase diagrams in Fig.~\ref{fig:clsr_phase} have completely different characteristics to those in Fig.~\ref{fig:plsr_phase}. The HM1 monolayer again shows no complete switching for $\mu_0 H_{\mathrm{ext}}=0.5$\,T in the whole phase space. For a 0.3\,T larger field the case is different. The switching probability at high temperatures reaches a maximum in the range of $x=-5$\,nm to $x=-10$\,nm. This means that the magnetic field pulse must point in the writing direction during cooling below the Curie temperature (when the heat spot moves away from the bit). The temperature must be large enough that the thermal reduction of the switching field is sufficient large for the given external field to reverse the magnetization of the HM1 material. After the field has changed its state the temperature of the heat pulse must be low enough to prevent the grain from being once again reversed (in the wrong direction). 

The phase diagram of the HM1 monolayer and $\mu_0 H_{\mathrm{ext}}=0.8$\,T (Fig.~\ref{fig:clsr_phase}a) displays narrow transition areas between complete and no switching of the grain. These transitions are larger for the HM2/SM grain, which favors adjacent track erasure. The core with complete switching is more narrow. In contrast to the HM1 grain, the HM2/SM grain also shows 100\,\% switching in a small parameter range in case of a smaller external field of 0.5\,T. 

The effect of the head velocity on the switching probability is illustrated in Fig.~\ref{fig:clsr_vel_comp}. As expected, higher head velocities enlarge the areas with complete switching and the phase transition in down-track direction. With a fixed FWHM of the heat spot and external field duration the switching probability in down-track direction $x$ is just scaled with the head velocity. The down-track width of the complete switching core is about $v_{\mathrm{head}}t_{H_{\mathrm{ext}}}$. The phase diagrams can be seen as footprints of the recording head. At high peak temperatures, above $T_{\mathrm{C}}$, the switching behavior of the HM1 grain becomes worse for increasing $v_{\mathrm{head}}$. This is also not surprising, because the duration of the heat pulse at the grains becomes shorter for higher velocities. If the pulse duration becomes comparable to the time the bits need for magnetization reversal, the switching probability decreases. This decrease starts at very high temperatures. The reason is that the effective time window for reversal starts with the cool down below the Curie point. For high peak temperatures the thermal gradient increases and the effective time window slightly below $T_{\mathrm{C}}$ narrows. The same effect was observed for PLSR (see Fig.~\ref{fig:plsr_phase}a) or for granular media in \cite{zhu2013understanding,zhu2015medium}.

\subsubsection{bit error rate}
\label{sec:BER_CLSR}
With the switching phase diagrams in Figs.~\ref{fig:clsr_phase} and \ref{fig:clsr_vel_comp} the BER according to Eq.~\ref{eq:BER} can be computed for various grain spacings, write temperatures and heat spot positions, as was done in Sec.~\ref{sec:BER_PLSR}. Here, $l_x$, $l_y$ and $T_{\mathrm{write}}$ cannot be optimized at the same time, because the down-track spacing is fixed with the choice of the head velocity. To achieve high AD we constrain the optimizations to a head velocities of 7.5\,m/s and 10\,m/s yielding $l_x=7.5$\,nm and $l_x=10$\,nm (because of $t_{H_{\mathrm{ext}}}=1$\,ns). Obviously, the HM1 monolayer has the smallest transition jitter values, and thus is the most promising candidate for high AD. The results of the HM1 monolayer were already published in \cite{vogler_heat_2015}. In this study the main objective is to throw light on the basic mechanisms of HAMR, to work out the benefits and disadvantages of different techniques and structures. Hence, the HM2/SM structure is investigated in more detail in the following.

\begin{table}[h!]
  \centering
  \vspace{0.5cm}
  \begin{tabular}{c c c c c c}
    \toprule
    \toprule
      & $\sigma_{\mathrm{T}}$\,[$\%\,T_{\mathrm{C}}$] & $l_x$\,[nm] & $l_y$\,[nm] &  $T_{\mathrm{write}}$\,[K]& AD\,[Tb/in$^2$] \\
    \midrule
      \multicolumn{6}{c}{HM2/SM grain}\\
    \midrule
      S & 3.0 & 7.5 & 14.0 & 720.0 & 4.61\\
      C & 3.0 & 7.5 & 19.5 & 705.0 & 3.31\\
      S & 3.0 & 10.0 & 10.5 & 715.0 & 6.14\\
      C & 3.0 & 10.0 & 18.0 & 625.0 & 3.58\\
    \midrule
      S & 3.7 & 10.0 & 11.5 & 715.0 & 5.61\\
      C & 3.7 & 10.0 & 20.05 & 700.0 & 3.15\\
    \midrule
      \multicolumn{6}{c}{HM1 grain*}\\
    \midrule
      SR & 3.15 & 7.5 & 6.5 & 715.0 & 13.23\\
      CR & 3.15 & 7.5 & 13.0 & 595.0 & 6.62\\  
    \bottomrule
    \bottomrule
  \end{tabular}
  \caption{\small Optimal center to center grain spacings, write temperatures and areal storage densities for shingled (S) and conventional (C) CLSR of HM2/SM bilayers with and without considering magnetostatic interactions. $\mu_0 H_{\mathrm{ext}}$ is assumed to be 0.8\,T. The shown values of $l_y$, $T_{\mathrm{write}}$ and the AD are all outputs of the optimization process.\newline
  *The results for the pure HM1 grain were already published in \cite{vogler_heat_2015}.}
  \label{tab:clsr_BER}
\end{table}
Table~\ref{tab:clsr_BER} summarizes the results of the AD optimizations. If magnetostatic interactions are neglected ($\sigma_{\mathrm{T}}=\sigma_{T_{\mathrm{C}}}$) the maximum AD is 6.14\,Tb/in$^2$ for shingled CLSR of the HM2/SM composite structure. Interestingly, a symmetric pattern with similar down-track and off-track spacings achieves a higher density than an asymmetric pattern with small $l_x$ and large $l_y$. For PLSR it is obvious that a symmetric pattern is beneficial, because the temperature decreases likewise in both directions. For CLSR the down-track spacing is determined by the head velocity and not by a temperature decrease. By looking at Fig.~\ref{fig:clsr_vel_comp} the reason becomes clear. The core of 100\,\% switching is very narrow for $l_x=7.5$\,nm. Hence, the transition jitter in off-track direction (peak temperature axis) becomes very large at low peak temperatures. Since a distribution of the Curie temperature deteriorates the jitter even more, only large off-track spacings are possible. With magnetostatic interactions our optimizations do not even yield any parameters with $\text{BER}<10^{-3}$, because the switching core is almost vanishing. A head velocity of 10\,m/s yields a maximum AD of 5.61\,Tb/in$^2$ for shingled CLSR with symmetric grain spacings and 3.15\,Tb/in$^2$ for conventional CLSR with an aspect ration of 1:2. 

The HM1 grain shows the same aspect ratios. Due to the much smaller switching probability phase transitions significantly higher AD beyond 10\,Tb/in$^2$ are obtained \cite{vogler_heat_2015}. 

\begin{figure}
 \includegraphics{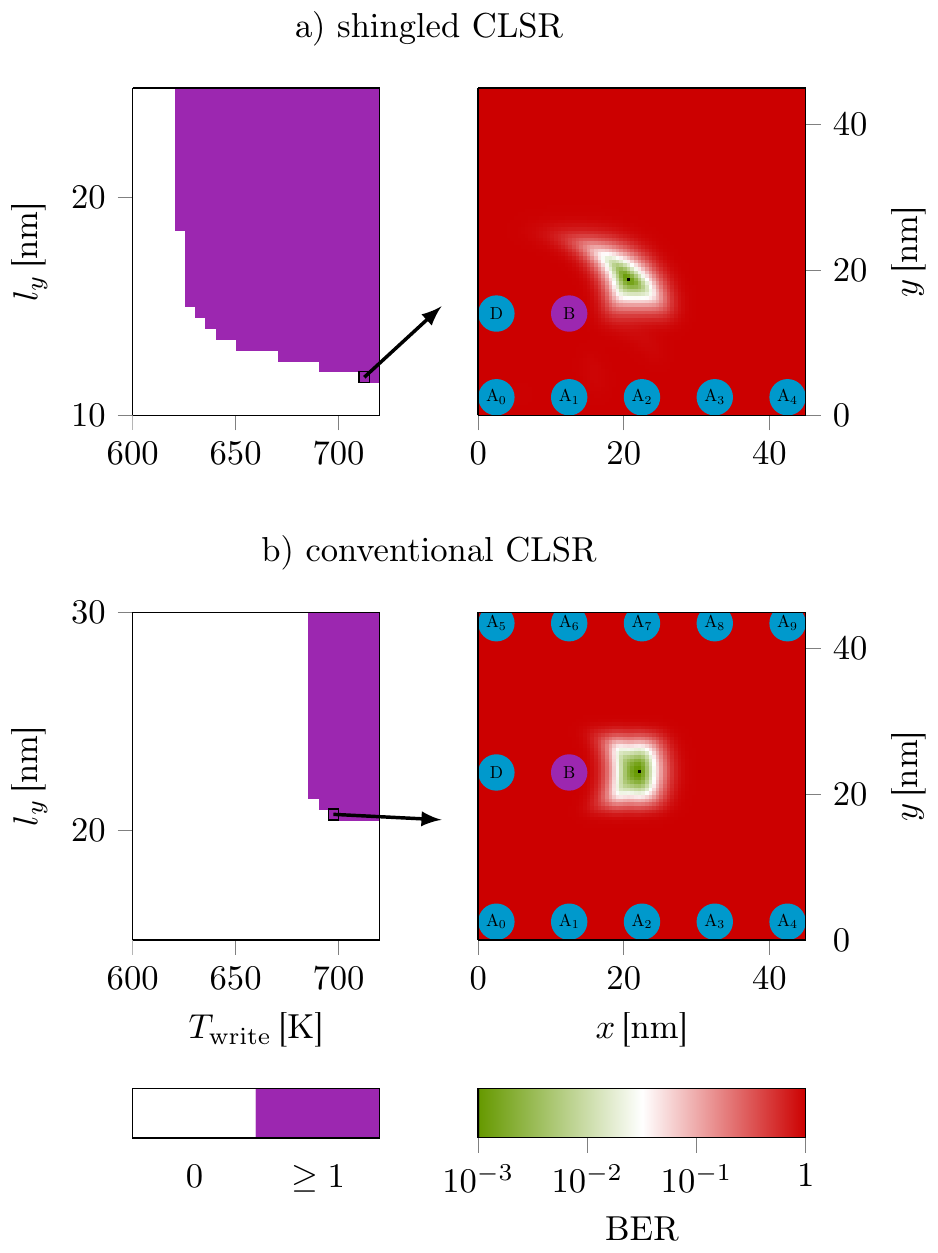}
\caption{\small (color online) AD optimizations for a) shingled and b) conventional CLSR of a medium consisting of HM2/SM grains, subject to a write field of 0.8\,T.\newline
(left) On the left a map of possible off-track spacings $l_y$ and write temperatures $T_{\mathrm{write}}$ to successfully switch bit B under the constraint of $\text{BER}<10^{-3}$ is shown. A distribution of the Curie temperature of 3.7\,\% and a displacement jitter according to Eqs.~\ref{eq:displacement_jitter} and \ref{eq:displacement_jitter_tot} are considered.\newline
(right) In the case of the setup with the highest possible AD, the BER for various heat spot positions $x$ and $y$ is shown in detail on the right. The color code is the same as in Fig.~\ref{fig:plsr_BER}.}
  \label{fig:clsr_BER}
\end{figure}
Figure~\ref{fig:clsr_BER} illustrates the AD optimization process and BER maps for the case of maximum AD for the HM2/SM grain, an external field of 0.8\,T and a distribution of the Curie temperature of $\sigma_{\mathrm{T}}=3.7$\,\%. As for PLSR $l_y$ increases for increasing $T_{\mathrm{write}}$ in case of shingled recording. The off-track spacing is again almost independent from the write temperature, due to the large transition jitter and the requirement of $n=1000$ (as discussed in Sec.~\ref{sec:BER_PLSR}) for conventional CLSR. In Ref.~\cite{vogler_heat_2015} a decreasing off-track spacing for decreasing write temperatures was reported for conventional CLSR of pure HM1 grains. This is caused by the much smaller transition jitter in off-track direction. Hence, write temperatures slightly above the transition are in preference to very high ones. For the same reason the optimal write temperature is 625\,K for conventional CLSR and $\sigma_{\mathrm{T}}=3.0$\,\% (see Tab.~\ref{tab:clsr_BER}).  

\subsubsection{transition jitter}
\label{subsubsec:jitter_CLSR}
In case of CLSR two different transition jitters, in down-track and in off-track direction, must be considered. Along the peak temperature axis (off-track direction) the transition jitter is calculated as explained in Sec.~\ref{subsubsec:jitter_PLSR}. Since the switching phase diagrams display a relation between the switching probability and the distance in down-track direction the curves for fixed $x$ can be directly fitted with the cumulative distribution function of a standard normal distribution. 
\begin{table*}
  \centering
  \vspace{0.5cm}
  \begin{tabular}{c c c c c c c}
    \toprule
    \toprule
      grain & $\sigma_{\mathrm{T}}$\,[$\%\,T_{\mathrm{C}}$] & $T_{\mathrm{write}}$\,[K] & $v_{\mathrm{head}}$\,[m/s] & $\sigma_{dP/d\mathrm{D},\mathrm{off}}$\,[nm] & $\sigma_{dP/d\mathrm{D},\mathrm{down}}$\,[nm] \\
    \midrule
      HM2/SM & 0.0 & 620.0 & 7.5 & 0.92 & 0.91\\
      HM2/SM & 0.0 & 620.0 & 10.0 & 1.15 & 1.00\\
      HM2/SM & 0.0 & 720.0 & 7.5 & 0.71 & 0.74\\
      HM2/SM & 0.0 & 720.0 & 10.0 & 0.81 & 0.82\\
      HM1 & 0.0 & 620.0 & 7.5 & 0.35 & 0.26\\
      HM1 & 0.0 & 620.0 & 10.0 & 0.40 & 0.27\\
      HM1 & 0.0 & 720.0 & 7.5 & 0.25 & 0.22\\
      HM1 & 0.0 & 720.0 & 10.0 & 0.26 & 0.25\\
    \midrule
      HM2/SM & 3.0 & 620.0 & 7.5 & 1.31 & 1.04\\
      HM2/SM & 3.0 & 620.0 & 10.0 & 1.52 & 1.10\\
      HM2/SM & 3.0 & 720.0 & 7.5 & 1.01 & 0.82\\
      HM2/SM & 3.0 & 720.0 & 10.0 & 1.08 & 0.88\\
    \midrule
      HM1 & 3.15 & 620.0 & 7.5 & 0.82 & 0.55\\
      HM1 &  3.15 & 720.0 & 7.5 & 0.60 & 0.28\\
    \midrule
      HM2/SM & 3.7 & 620.0 & 10.0 & 1.64 & 1.17\\
      HM2/SM & 3.7 & 720.0 & 10.0 & 1.20 & 0.92\\
    \midrule
      HM2/SM & 4.25 & 620.0 & 7.5 & 1.60 & 1.17\\
      HM2/SM &  4.25 & 720.0 & 7.5 & 1.24 & 0.90\\
    \bottomrule
    \bottomrule
  \end{tabular}
  \caption{\small Transition jitters in off-track and down-track direction for various Curie temperature distributions, head velocities and write temperatures, in the case of a HM2/SM and a HM1 grain subject to an external magnetic field of 0.8\,T.}
  \label{tab:clsr_transition_jitters}
\end{table*}
In Tab.~\ref{tab:clsr_transition_jitters} transition jitters in both directions and for both considered grains (HM1 and HM2/SM) are illustrated for various distributions of the Curie temperature and various head field velocities. If magnetostatic interactions are considered only transition jitters for parameters leading to maximum AD are evaluated.

As expected the transition jitters are larger for lower write temperatures. This holds for both grains and independent from the head velocity and $T_{\mathrm{C}}$ distribution. Without any distribution of the Curie temperature ($\sigma_{\mathrm{T}}=0$\,\%) the jitters in down-track and off-track direction are very similar. For larger $T_{\mathrm{C}}$ distribution the jitter in down-track direction does not increase very much, in contrast to $\sigma_{dP/d\mathrm{D},\mathrm{off}}$. This is not surprising, because the Curie temperature distribution directly affects only the switching probabilities along the peak temperature axis. The broadening in down-track direction is an indirect effect. If one compares Tabs.~\ref{tab:plsr_transition_jitters} and \ref{tab:clsr_transition_jitters} one notices that the transition jitters in off-track direction for CLSR and those for PLSR in case of the HM2/SM grain are almost identical. The values are throughout a bit better for a head velocity of 7.5\,m/s and a bit worse for 10\,ms. The lower optimal down-track spacings, and thus higher AD, for shingled CLSR can be explained with the lower down-track transition jitter if $T_{\mathrm{C}}$ distributions are considered. 

In case of the pure HM1 grain the transition jitters are significantly lower in both directions. If magnetostatic interactions are considered the jitter is still smaller than for the HM2/SM grain without any $T_{\mathrm{C}}$ distributions. Additionally the strayfield does not shift the phase transition much, resulting in a low $\sigma_{\mathrm{T}}$, and thus a small jitter. These are clearly the reasons for the high AD found in \cite{vogler_heat_2015}.

Compared to magnetic recording of exchange spring-media without heat-assist \cite{suess_superior_2015} one has to mention that in case of HAMR the transition jitters for the HM2/SM grain with magnetostatic interactions are larger in off-track direction and comparable in down-track direction. The HM1 grain shows a significant lowering of the transition jitter in both directions.

\section{Conclusion and Outlook}
\label{sec:conclusion}
To conclude, with a coarse grained LLB model for exchange coupled Curie temperature modulated grains, postulated in a previous work \cite{volger_llb}, we extensively examined the switching behavior of small cylindrical recording grains ($d=5$\,nm, $h=10$\,nm) during HAMR of bit-patterned media. In detail we considered two different recording techniques, one where the heat spot is pulsed (PLSR) and one where a continuous spot moves over the medium (CLSR). In addition shingled and conventional recording were discussed. With the computed switching phase diagrams bit error rates (BER) for various heat spot positions, grain spacings and write temperatures could be determined. To picture the whole write process as realistic as possible we assumed a distribution of the grain's Curie temperatures, a displacement jitter of the head and the bit positions and a size distribution of the bits according to the ASTC guidelines. We investigated two grain types, a pure hard magnetic one, which has similar properties as FePt and a bilayer structure with graded Curie temperature comparable to Fe/FePt, where Fe is for example doped with Tb to achieve high damping.

In case of PLSR the maximum achieved areal densities (AD) were slightly below 5\,Tb/in$^2$ for the bilayer structure. The main problem is that due to the short heat pulse the switching of the grains is not sufficiently reliable with 0.8\,T external field, even if grains with graded Curie temperature are used to further assist the write process. Hence, the transition jitter of the switching probability is too large to reach higher AD. For CLSR of the same bilayer structure the maximum AD is a bit higher than for PLSR. Although, the transition jitter in off-track direction is similar to the calculated PLSR value, the according jitter in down-track direction is clearly smaller, yielding smaller down-track spacing, and thus a higher AD with 5.61\,Tb/in$^2$. Because of the high saturation magnetizations of Fe and FePt the performance of the graded structure suffers from large magnetostatic interactions involving increased transition jitters. 

The pure hard magnetic grain displays very narrow transitions. The obtained jitters are significantly lower than for magnetic recording of exchange coupled media without heat-assist. Hence, CLSR with an AD beyond 10\,Tb/in$^2$ was reported in \cite{vogler_heat_2015}. For the given magnetic write fields PLSR of the single phase grain does not reach a 100\,\% switching probability at elevated temperatures. The reason are again the short heat pulses yielding a lack of reversal time for the magnetic moments, and thus short effective recording time windows.

Note, the presented promising AD are based on bit-pattered media with very narrow size distributions. Manufacturing such patterns is not yet feasible, but is expected to be in the future. Additionally successfully written patterns have to be read with a low signal to noise ratio. Although a reader design for bit-pattered media with a density of 10\,Tb/in$^2$ was proposed in \cite{wang_reader_2013} a lot of work on the reader side has to be done until a working prototype with a user density of 10\,Tb/in$^2$ can be realized.

Altogether, our simulations suggest that shingled CLSR is the most promising HAMR technique for future high density bit-patterned media. Multilayer structures with graded Curie temperatures and optimized saturation magnetizations \cite{suess_exchange_2005} could improve the switching performance of these grains to possibly outreach the obtained AD for hard magnetic single phase grains. To further decrease the thermal jitter multilayer structures like Fe/FeRh/FePt with an interlayer showing a first order phase transition, to switch the interactions between the soft and hard magnetic parts of the grain on and off, are also thinkable.

\section{Acknowledgements}
The authors would like to thank the Vienna Science and Technology Fund (WWTF) under grant MA14-044 and the Austrian Science Fund (FWF): F4102 SFB ViCoM, for financial support. The support from the CD-laboratory AMSEN (financed by the Austrian Federal Ministry of Economy, Family and Youth, the National Foundation for Research, Technology and Development) was acknowledged. The computational results presented have been achieved using the Vienna Scientific Cluster (VSC).

%

\end{document}